\magnification 1200

\centerline{CONFORMAL FIELD THEORY AND GRAPHS $^{*}$}
\footnote{}{${}^{*)}$~ Talk given by V.B.P. at the 21st International
Colloquium on Group Theoretical Methods in Physics, Goslar,
Germany, 15-20 July 1996.}
\bigskip

\centerline{V.B. Petkova$^{1}$ and J.-B. Zuber$^{2}$}

\centerline{${}^{1}$Institute for Nuclear Research
and Nuclear Energy,}
\centerline{ Tsarigradsko Chaussee 72, 1784 Sofia, Bulgaria}
\medskip

\centerline{${}^{2}$CEA--Saclay, Service de Physique Th\'eorique}
\centerline{F-91191 Gif sur Yvette Cedex, France}
\bigskip
\bigskip


\def\za{\alpha}  \def\zg{\gamma} \def\zd{\delta}
\def\ze{\varepsilon}    
\def\zl{\lambda} \def\zm{\mu} \def\zn{\nu} \def\zo{\omega}
 \def\zr{\rho}

\def\GLh{\hat{\Lambda}}

\def\CP{{\cal P}}\def\CV{{\cal V}}
\def\CN{{\cal N}}\def\CB{{\cal B}}
\def\IZ{Z\!\!\!Z}\def\CG{{\cal G}}\def\CU{{\cal U}}

\def\un{{\bf 1}}
\def\Exp{{\rm Exp}}

\def\slh{\widehat{sl}}
\def\av{a^{\vee}}
\def\bv{b^{\vee}}

\input epsf.tex
\newcount\figno
\figno=0
\def\fig#1#2#3{
\par\begingroup\parindent=0pt\leftskip=1cm\rightskip=1cm\parindent=0pt
\baselineskip=11pt
\global\advance\figno by 1
\midinsert
\epsfxsize=#3
\centerline{\epsfbox{#2}}
\vskip 12pt
{\bf Fig. \the\figno:} #1\par
\endinsert\endgroup\par}
\def\figlabel#1{\xdef#1{\the\figno}}
\def\encadremath#1{\vbox{\hrule\hbox{\vrule\kern8pt\vbox{\kern8pt
\hbox{$\displaystyle #1$}\kern8pt}
\kern8pt\vrule}\hrule}}

The ADE graphs -- the root diagrams of the simply laced simple
Lie algebras appeared in the framework of the
two-dimensional conformal field theories (CFT) based on the affine
algebra $\slh{(2)}_k$
in the   classification of modular invariant
partition functions [1] and in some related lattice
models [2]. The modular invariants are labelled by the
Coxeter number $h=k+2$ while  their diagonal terms, encoding the 
scalar field content of the theory, are enumerated by the
Coxeter exponents. Graphs describing the spectrum of some
invariants  for the $\slh{(n)}$ theories were proposed in
[3] ($\IZ_n -$ orbifolds) and in [4] (some cases of
$\slh{(3)}$).  The aim of the reviewed work 
[5] is to  further extend these 
results exploiting some deeper relations with structures in CFT.

We first recall some basic facts.
The WZNW genus one partition functions are sesquilinear forms on
the affine algebra characters with integer coefficients
$\CN_{\zl\,\bar\zl}\,,$ 
$$
Z^{(n+k)}(q)= \sum_{\zl,\bar\zl } \CN_{\zl\,\bar\zl}\,
\chi_{_{\scriptstyle 
\zl}}(q)\, \chi_{_{\scriptstyle \bar\zl}}(\bar q)\,, \qquad 
\zl\,,\,\bar{\zl} \in \CP_{++}^{(k+n)}\,, \quad
\CN_{\zr\,\zr}=1\,. 
\eqno(1)$$
Here $\CP_{++}^{(k+n)}$   is set of integrable heighest weights
 of the affine algebra
$\slh(n)_k$ (shifted by  $\rho=(1,1,\dots,1)$). 
The Type I theories are those for which there exists
an algebra extending the chiral algebra and the
partition function  may be recast in a block-diagonal form
$$
 Z^{(n+k)}= 
\sum_i\, |\sum_{\zl\in \CP_{++}^{(k+n)}}\,
{\rm mult}_{\CB_i}(\zl)\,  \chi_{_{\scriptstyle
\zl}} |^2 =\sum_{ \CB_i}| \chi_{_{\scriptstyle \CB_i}} |^2\,,
\eqno(2)$$
where $\CB_i$ ($\chi_{_{\scriptstyle\CB_i}}$ ) are
representations (characters) 
of the extended chiral algebra and\hfil\break
 ${\rm mult}_{\CB_i}(\zl)\,$
is the multiplicity of $\zl\in \CP_{++}^{(k+n)}\,$ in $\CB_i$.
The modular invariance of 
(2) implies the relation
$$
\sum_{\zo \in {\CP}^{(h)}_{++}}\, {\rm mult}_{\CB_i}(\zo)\,
S_{\zg \zo} =\sum_{\CB_j}{\rm mult}_{\CB_j}(\zg)\,  S_{\CB_j\,
\CB_i} \,,
\eqno(3)$$
where $S_{\zg \zo}$ and $  S_{\CB_j\, \CB_i}$ are the initial and
extended modular matrices; mult$_{\CB_i}(\zr)=\zd_{\CB_i \un}.$

We postulate that
each of the looked for graphs $\CG$ satisfies  a set of
requirements: namely it
is connected and 
unoriented, i.e., described by an irreducible, symmetric
adjacency matrix 
${}^tG =G$ with $G_{ab}$--nonnegative integers;
in  the set of vertices $\CV\,$ 
 a $\,\IZ/n\IZ$ grading $a \mapsto \tau(a)$, the ``$n$-ality",
is introduced, so that $G_{ab}\not = 0$ only if $\tau(a)\not
=\tau(b)$. 
This enables one to
split  this adjacency matrix into a sum of $n-1$ matrices
$G=G_1+G_2+\cdots +G_{n-1}\,,$
where 
$ (G_p)_{ab}\ne 0 \qquad{\rm only \ if \ }\quad \tau(b)=\tau(a)+p
\ {\rm  mod } \,\, n\,,$ and $\,{}^tG_p=G_{n-p}\,.$
Accordingly, the graph may be regarded as the superposition on the
same set of vertices of $n-1$
oriented (except for $p=n/2\,$),  not   all necessarily
connected, 
 graphs $\CG_p$
of adjacency matrices $G_p$, $p=1,\cdots, n-1$.
We will furthermore
 require that  there exists an involution $a\mapsto
\av$ on $\CV$ 
such that $\tau(\av)=-\tau(a)$ and
$(G_p)_{ab} = (G_p)_{\bv\av}\ .$
 The matrices $G_p$ are assumed to commute among themselves, 
hence are ``normal",
i.e. simultaneously  diagonalisable in an orthonormal basis;
  the common eigenvectors are
labelled by integrable weights $\zl\in \CP_{++}^{(k+n)}$ for some
level $k$, we  denote them $\psi^{\zl}=( \psi^{(\zl)}_a)\,, a\in
\CV$. 
The set of these $\zl$, some of which
may occur with multiplicities larger than one, will be denoted by
$\Exp$.
We require that
 the  eigenvalues $\zg_p^{(\zl)}$ of $G_p$
coincide with the corresponding  eigenvalues of the
Verlinde fundamental matrices 
$N_{\GLh_p+\zr}\,,$ i.e., 
$\zg_p^{(\zl)}= {S_{\GLh_p+\zr\, \zl}/
 S_{\zr\, \zl}}\,,$ where 
 $\GLh_1,\cdots,\GLh_{n-1}$ are the fundamental weights.
It is assumed that the identity representation
  $\zr    \in \Exp \,$  and that it appears with
multiplicity one: it corresponds to the eigenvector of largest
eigenvalue of $G_1\,,$  $\gamma^{(\zr)}_{1}\ge
|\gamma^{(\zl)}_{1}|\,,$ the 
so--called Perron--Frobenius eigenvector;
 its components  $\psi_a^{(\zr)}\,,$ $a\in \CV\,,$ are
positive. 
The eigenvector matrices $\psi_a^{(\zl)}\,$ 
replace the (symmetric)
 modular matrices
$S_{\zm \zl}$ of the diagonal theory
and satisfy the relations
$$
\sum_{a\in \CV} \  \psi^{(\zl)}_a\,\psi^{(\zn)\, *}_a
= \zd_{\zl \zn}\,, \quad
\sum_{\zl \in \Exp }\  \psi^{(\zl)}_a\, \psi^{(\zl)\, *}_c =
\zd_{a c}\,,
\qquad \psi^{(\zl)}_{a^{\vee}} =  \psi^{(\zl)\, *}_a=
\psi^{(\zl^*)}_a\,.
$$
It is assumed that there is at least one vertex denoted by $\un
\,, \, \un^{\vee}=\un\,,\, \tau(\un)=0\,, $ such that
$\psi_{\un}^{\lambda}>0$ for all $\zl \in \Exp$.

Accordingly
one  introduces
two sets of real numbers providing two extensions of the Verlinde
 formula for the fusion rule multiplicities,
$$
N_{ab}^{\,\, c}=
\sum_{\zl \in \Exp } \ {\psi^{(\zl)}_a\psi^{(\zl)}_b\psi^{(\zl)\,
*}_c \over \psi^{(\zl)}_{\un}} \,, \qquad
 M_{\zl \zm}^{\ \ \zn}=
\sum_{a\in \CV} \  {\psi^{(\zl)}_a \,\psi^{(\zm)}_a\,\psi^{(\zn)\,
*}_a \over \psi^{(\zr)}_a}\,.
\eqno(4)$$
These two sets of real numbers (assumed furthermore to be
nonnegative 
for the Type I theories) can be looked as the structure constants
of a pair of associative, commutative algebras $\cal U $
and $\hat{\cal U} $, with identity and
involution, 
and with $|\CV|=|\Exp|$ one-dimensional representations provided
by the eigenvalues, 
i.e., as a dual pair of $"C-$algebras", [6]. 
In particular in the diagonal cases
the two algebras are selfdual and
coincide, the diagonalisation matrix $\psi_a^{(\zl)}\,$ being
replaced 
by the symmetric matrix $S_{\zm \zl}$.  The relevance
of the $C-$algebras with nonnegative structure
constants for the study of Type
I theories was first pointed out and exploited in [4].
There is another generalisation of the Verlinde  formula, also
studied extensively in [4], namely the set of integers
defined according to
$$
 V_{\zg a}^b = \sum_{\zo \in \Exp}\,
{S_{\zg \zo}
\over S_{\zr \zo} }\,\psi^{(\zo)}_a\, \psi^{(\zo)\, *}_b\,, 
\qquad a,b \in \CV\,, \ \ \zg\in 
 {\CP}^{(k+n)}_{++} \,,
\eqno(5)$$
in particular $ V_{\GLh_p +\zr}=G_p$.
The matrices $V_{\zl}$ realise a  representation of the Verlinde
fusion 
algebra,
while 
$V_b$ for any $b\in \CV$,  considered as a rectangular
matrix 
$(V_{b})_{\zl}^c=V_{\zl b}^c\,,$ intertwines the adjacency
matrices 
of the  diagonal and the nondiagonal graphs for a given
value of the level $k$, i.e., 
$N_{\GLh_p+\zr}\, V_b= V_b\,G_p\,.$
Note also the relation
$$
 V_{\zg } \, N_a =\sum_{b\in \CV} \,  V_{\zg a}^b\, N_b\,, \ \
{\rm or}\,,\quad  V_{\zg }  =\sum_{b\in \CV} \,  V_{\zg \un}^b\,
N_b\,, 
\eqno(6)$$
which implies that the adjacency matrices $G_p$ belong to the 
$N$ algebra being expressed as linear combinations 
(with nonnegative integer coefficients) of the basis $\{N_a\}$.

Along with a few exceptional cases
there are (restricting to embeddings into simple algebras
$\hat{g}_1$) four 
infinite series of conformal embeddings
of the algebra $\slh{(n)}$ [7]
$$
\slh{(n)}_{n-2} \subset   \slh{\Big({n(n-1)\over
2}\Big)}_{1}\,, \qquad n \ge 4\,, 
$$
$$
\slh{(n)}_{n+2}  \subset \slh{\Big( { n(n+1)\over
2}\Big)}_1\,, 
\eqno(7)$$
$$
\slh{(2n+1)}_{2n+1} \subset  \widehat{so}(4n(n+1))_1\,,
$$
$$
\slh{(2n)}_{2n} \subset  \widehat{so}(4n^2-1)_1\,, \qquad
n \ge 2\,. 
$$

The set $\Exp$ splits into classes $B_i$ -- the integrable
representations of $\hat{g}_1$. We shall also  denote
the elements in $\Exp$ by
$(\zl,i,\ze_i)$, $\zl\in \CB_i\,,$ $|\ze_i| =$ mult$_{\CB_i}(\zl)$,
to distinguish $\zl\in {\CP}^{(k+n)}_{++}$ appearing in different
extended algebra representations as well as with a nontrivial
multiplicity within a given extended representation;
for simplicity of notation the indices $i,\ze$ are sometimes
omitted.  The physical fields are labelled by pairs $\Big((\zl,
i, \ze_i)\,, \,(\bar{\zl}, i,\bar{\ze}_i)\Big)$, $\zl, \bar \zl
\in \CB_i$. 

 It was observed in [5] that in the case
 $sl{(2)}$  the nondiagonal solutions for
the  relative  structure constants accounting for the
contribution of the scalar fields in the expansion of products of
scalar physical operators coincide 
with the set of algebraic numbers $M_{\zl \zm}^{\zg}$ in
(4) for the corresponding $D-$, or $E-$ type graph.
Furthermore in the cases described by conformal embeddings the
same numbers determine uniquely the more general spin fields
structure constants, since the (square of the) latter factorise
into a left and right chiral parts precisely given by a pair of
$M-$ algebra structure constants.  This property of the models
related to conformal embeddings allows to block-diagonalise the
duality equations for the relative structure constants thus
recovering the  extended model fusion (crossing) matrices
appearing in the corresponding diagonal equations.  
In a more restricted version this idea has been worked out
and extended for the higher rank cases $\slh{(n)}$ in [5]
assuming that in all theories described by conformal embeddings a
kind of chiral factorisation of the  general structure constants
takes place.  This led us to the following set of equations for
the chiral pieces denoted $M_{\zl \zm}^{\zg}$ by analogy with the
$n=2$ case 
$$
N_{\CB_i\, \CB_k}^{\CB_j}  =\sqrt{D_{\CB_i}
\over D_{\lambda} } \,
\sqrt{D_{\CB_k} \over D_{\mu} }\,
\sum_{\zg \in
 {\CP}^{(k+n)}_{++}\,; \,  \ze_j }\, M_{(\zl,i,\ze_i)\,
(\zm,k,\ze_k) }^{(\zg,j,\ze_j)}\, 
\sqrt{ D_{\zg}\over D_{\CB_j}} \,,
\qquad \forall \, \, 
\zl\in \CB_i\,, \,
\zm\in \CB_k\,.
\eqno(8)$$
Here in the l.h.s.
$N_{\CB_i\, \CB_k}^{\CB_j} $ is the
extended 
Verlinde fusion rule multiplicity,  while the quantum dimensions
$D_{\lambda}$ 
and $D_{\CB_i}$ of the
initial and the extended theories appear in the r.h.s.; this data
is known for all of the conformal embeddings--in particular the
extended multiplicity in the l.h.s. of (8) takes only the
values 
$0,1$. The constants
$ M_{(\zl,i,\ze_i)\, (\zm,k,\ze_k)
}^{(\zg,j,\ze_j)}\equiv 0\,$ if 
the Verlinde multiplicies $N_{\zl\, \zm}^{\zg}\,,$ or 
$N_{\CB_i\, \CB_k}^{\CB_j}\, $
 vanish; the summation in the r.h.s. of
(8) is restricted within the class $\CB_j$.
In the case
$n=3$ the 
solutions of the set of algebraic equations (8) are
 consistent with the  values for the $M-$ algebra
matrix elements (4) found in [4] and furthermore
they  lead to new solutions for higher $n$, allowing to construct
the corresponding graphs, see [5] for explicit examples.

The set of equations (8) furthermore allows to determine
explicitly 
in general some of the
eigenvalues of the $M-$ matrices, i.e., some of the
$1-$ dimensional representations of the $M-$ algebra.  
To do that one establishes a one to one correspondence
between the set of integrable representations of the extended
Kac--Moody algebra $\hat{g}_1$ and a subset $T\subset \CV$ of the
vertices of the graph such that if $c\in T$ is identified with
a representation 
$\CB_i,$ then $c^{\vee}$ is also in $T$ and
is identified with 
$\CB_i^*$; we shall identify
the vertex  $\un \,$
with the  identity weight of the extended
algebra. One obtains then from  (8) an analytic formula
for ${\psi_{c}^{(\zl,i,\ze)}/ \psi_{c}^{(\zr)}}\,,$ for
$\forall \, c \in T\,,\, (\zl,i,\ze_i)\in \Exp\,,$
in terms of the extended modular matrices $S_{c\  B_i}\equiv
S_{B_j B_i}$ ($c\equiv B_j$), which furthermore implies
$$
{ \psi_{c}^{(\zl,i,\ze)} \over \psi_{\un}^{(\zl,i,\ze)}}=
{S_{c \, \CB_i }\,\over S_{\un \, \CB_i}}\,,
  \qquad \psi_{\un}^{(\zl, i, \ze)}
= \sqrt{S_{\zr\, \zl}\,  S_{\un\,  \CB_i}}\,,
\qquad  c \in T \,,  (\zl, i, \ze_i)\in \Exp\,.
\eqno(9)$$
According to (9) the (dual) Perron-Frobenius eigenvector
$\psi_{\un}$ has indeed positive components.  The formula
(9) provides general explicit expression for the subset of
components of the eigenvectors of the adjacency matrices
corresponding to the subset $T$. This allows to determine the
matrix elements $N_{a b}^c$ for $a,b,c \in T$ as coinciding with
the corresponding extended Verlinde matrix elements 
$N_{\CB_i\, \CB_k}^{\CB_j} $
for $a\equiv \CB_i\,, b\equiv \CB_k\,, c\equiv \CB_j\,.$
Furthermore assuming that both $N$ and $M$ structure 
constants are nonnegative one shows that the set $N_a\,, a\in
T\,,$ provides a subalgebra $\CU_T$ of the $N-$ algebra $\CU$
isomorphic to the extended Verlinde algebra $N_{\CB_i}$. The
set of vertices $\CV$ then splits into equivalence classes
$T_1=T, T_2,..., T_t\,$ which allows to define a $C-$ factor
algebra $\CU/\CU_T$ [6].  Alternatively the $M-$ algebra
$\hat{\cal U} $ has a $C-$ subalgebra $\hat{\cal U}_{\hat T}\,,$ 
described by the subset ${\hat T}$ of $\Exp$ appearing in the
decomposition of the identity representation 
of the
extended algebra, while the factor $C-$ algebra $\hat{\cal
U}/\hat{\cal U}_{\hat T}$ is isomorphic to the extended Verlinde
algebra. 
The relation (8)  then can be interpreted as
expressing the structure constants of   $\hat{\cal U}/\hat{\cal
U}_{\hat T}$ as an average of the structure constants of
$\hat{\cal U}$  over classes $\hat{T}_i$, i.e., a 
relation in the theory of $C-$ algebras with nonnegative
structure constants [6], realised here with explicit
specific values of the parameters, given by the quantum
dimensions.
Note that,
although obtained under assumptions which sound plausible only
for the cases of conformal embeddings, the relation  (8)
and its 
consequences presumably hold true also for the orbifold theories
(where one can determine explicitly the full eigenvector
matrices); 
we have checked this  for the $\IZ_n -$ orbifold graphs of [3],
see also the recent work [8] for
general expressions for the orbifolds extended modular matrices.

One further consequence of (8) is obtained
inserting  (9) into (5), namely
$$
  V_{\zg \un}^c =\sum_{\CB_i}\,
S_{\CB_j  \CB_i}^*
\sum_{\zo \in 
 {\CP}^{(k+n)}_{++} }\, {\rm mult}_{\CB_i}(\zo)\, 
S_{\zg \zo} ={\rm mult}_{\CB_j}(\zg)\,
, \qquad \CB_j\equiv c \in T\,,
\eqno(10)$$
using for the second equality in (10)
the consistency condition (3).
This determines the matrix $N_{\zl \bar \zl}$
in (1) in terms of the intertwiner $V_ {\un}\,,$
$$
\CN_{\zl \bar \zl}=\sum_{c\in T}\,   V_{\zl \un}^c\,   V_{\bar
\zl \un}^{c}\,, 
\eqno(11)$$
a property first empirically observed in [4], and
also  derived
recently  by Ocneanu [9] in a different context as
reflecting the counting of ``essential paths'' on the graph.

The set of vertices for which the components
$\psi_a^{(\zm)}$ 
are explicitly determined 
according to  (9) can be enlarged
beyond the subset $T$. Indeed  using (9) and (3)
we have
$$
\sum_{a\in \CV}\, (V_{\zl \un}^a)^2 =  \sum_{\za \in \un}\,
N_{\zl \zl^*}^{\za}\,,
\eqno(12)$$
where   $N_{\zl \zl^*}^{\za}$ are Verlinde multiplicities, and
hence 
$$
 \forall \zl\in  {\CP}^{(k+n)}_{++}\,, {\rm s. t.} \,
\sum_{\za \in \un}\, N_{\zl \zl^*}^{\za}=1\,, \quad   \exists \,\,
a_{\zl}\in \CV\,,  \, {\rm  s. t.}\,\,  V_{\zl \un}^a = \delta_{a
a_{\zl}}\,. 
\eqno(13)$$
Hence according to (6) $V_{\zl}\equiv N_{a_{\zl}}$.
This allows to determine $\psi_{a_{\zl}}^{(\zm)}$ using (9)
and the fact that $V_{\zl}$ admits eigenvalues identical to
eigenvalues 
of the Verlinde matrix $N_{\zl}$, c.f. (5). An example is
provided 
by the fundamental weight $\zl=\GLh_1+\zr$ for which the above
condition can be checked [10] to hold for all embeddings, so
that $(G_1)_{\un a}=\delta_{a a_{\GLh_1+\zr}}$ which implies that
the vertex $\un$ is ``extremal'', i.e., there is only one arrow
leaving it (or entering it), the corresponding vertex being
$a_{\GLh_1+\zr}$ (and $a_{\GLh_{n-1}+\zr}$). The latter property
of $\un$, assumed in [5], was proved in the recent paper
[10] starting from a more abstract setting. In fact this
property of $\un$  extends to all  
vertices $ a\in T$, such that $D_a=D_{\CB_i}=1$:
i.e., $(G_1)_{a b}=\delta_{a\, b(a)}\,$ for such $a$ and some
$b=b(a)$, and hence 
in the first three series in (7)
it extends to the full set $T$, 
thus allowing to recover the components $\psi_{b}^{(\zm)}$
corresponding to 
the class $T_{a_{\GLh_1+\zr}}\ni b$.

We expect that taking into account  properties as in the above
observations  suggested by the results in [5] and [10],
will make possible the construction of many new examples of graphs.
Let us add one such new example: consider
the embedding $\slh(4)_6 \subset \slh(10)_1$
(see [11] for the explicit expression for the modular 
invariant).
Denote the exponents $(\zl,i)\,, i=0,1,\dots,9\,,$ $|\Exp|=32.$
The graphs $\CG_p$ are recovered 
from the following
eigenvector matrix (with $\psi_{a_j}^{(\zl,i)}\,,\, a_j\in T\,,$
determined from (9), $a_0\equiv \un \equiv \CB_0\,, $ and 
$ S_{\CB_j\,  \CB_k} = {1\over \sqrt{10}}\, \exp{({2 \pi i \over 
10}\, j k)}$ )
$$
\psi_{b_j}^{(\zl,i)}:= \zg_1^{(\zl)}\,
\psi_{a_j}^{(\zl,i)}\,, \quad \psi_{c_j}^{(\zl,i)}:=
\zg_2^{(\zl)}\, \psi_{a_j}^{(\zl,i)}\,, \quad 
 j=0,1,\dots,9\,, \quad a_j\in T\,,\,
$$
$$
\psi_{d_0}^{(\zl,i)}:={S_{(2,2,1)\, \zl} \over S_{\zr \zl} } \, 
\psi_{\un}^{(\zl,i)} -\zg_1^{(\zl)}\,\psi_{a_9}^{(\zl,i)}\,,
\quad 
\psi_{d_5}^{(\zl,i)}:={S_{(5,2,2)\, \zl} \over S_{\zr \zl} } \,
\psi_{\un}^{(\zl,i)} -\zg_1^{(\zl)}\, \psi_{a_4}^{(\zl,i)}\,.
\eqno(14)$$
\fig{The graph of $G_1$ 
corresponding to the conformal embedding
$\slh(4)_6\subset \slh(10)_1$. The points of 4-ality 0 and 2
(resp 1 and 3)  are represented by open and black disks (resp
squares).  The involution $a\to a^\vee$
is the reflection in the horizontal diameter.
The vertices denoted by $a_j\,, b_j\,, c_j\,$ $(j=0,1,\dots
9)\,,$ in (14) lie clockwise on three (inward going) concentric
circles  respectively, while $d_0$ and $d_5$ denote the selected
(bottom and top respectively) vertices in the centre.}
{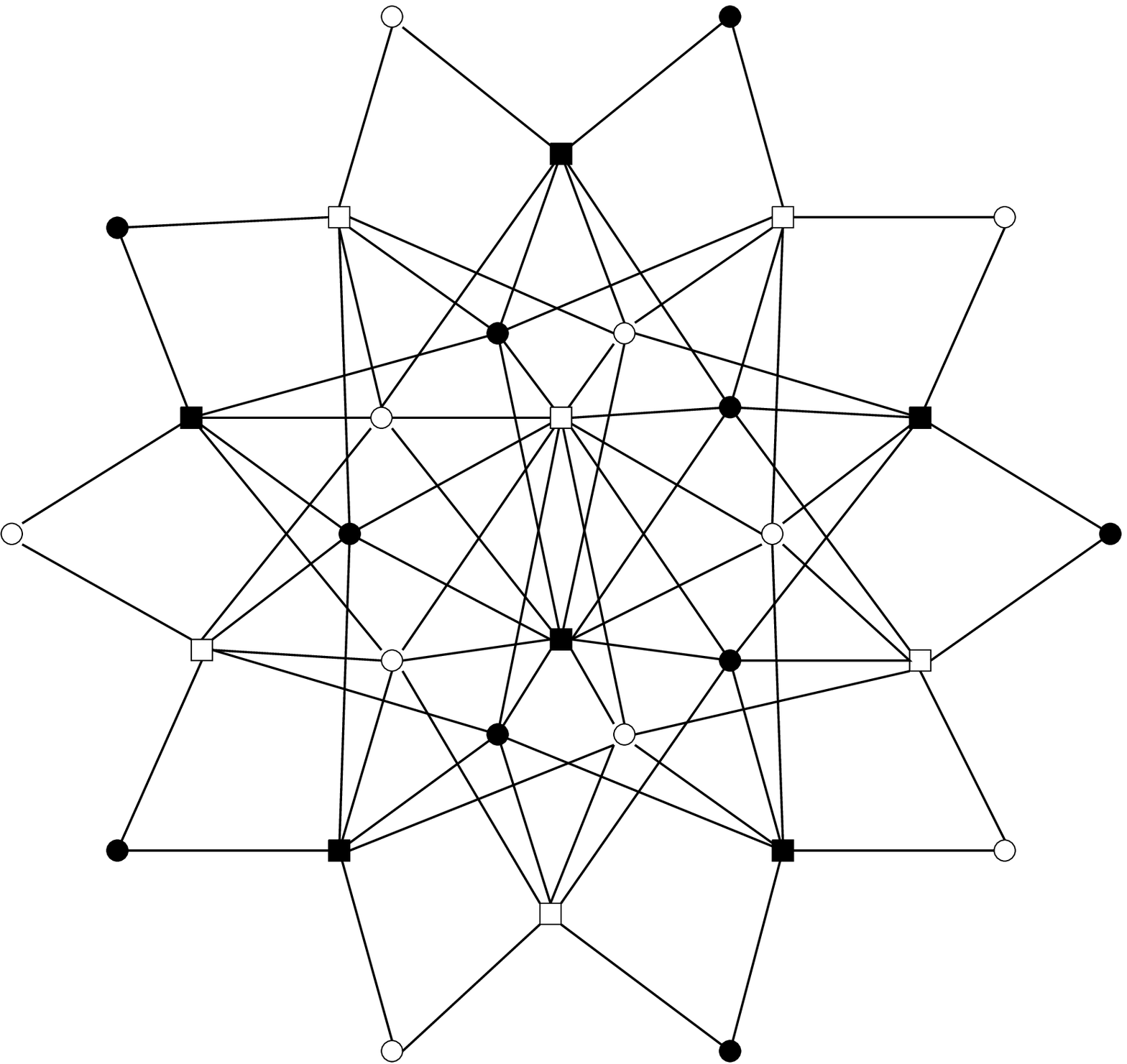}{64mm}
\figlabel\geeone

\fig{Graph of $G_2$ for the same
case.}{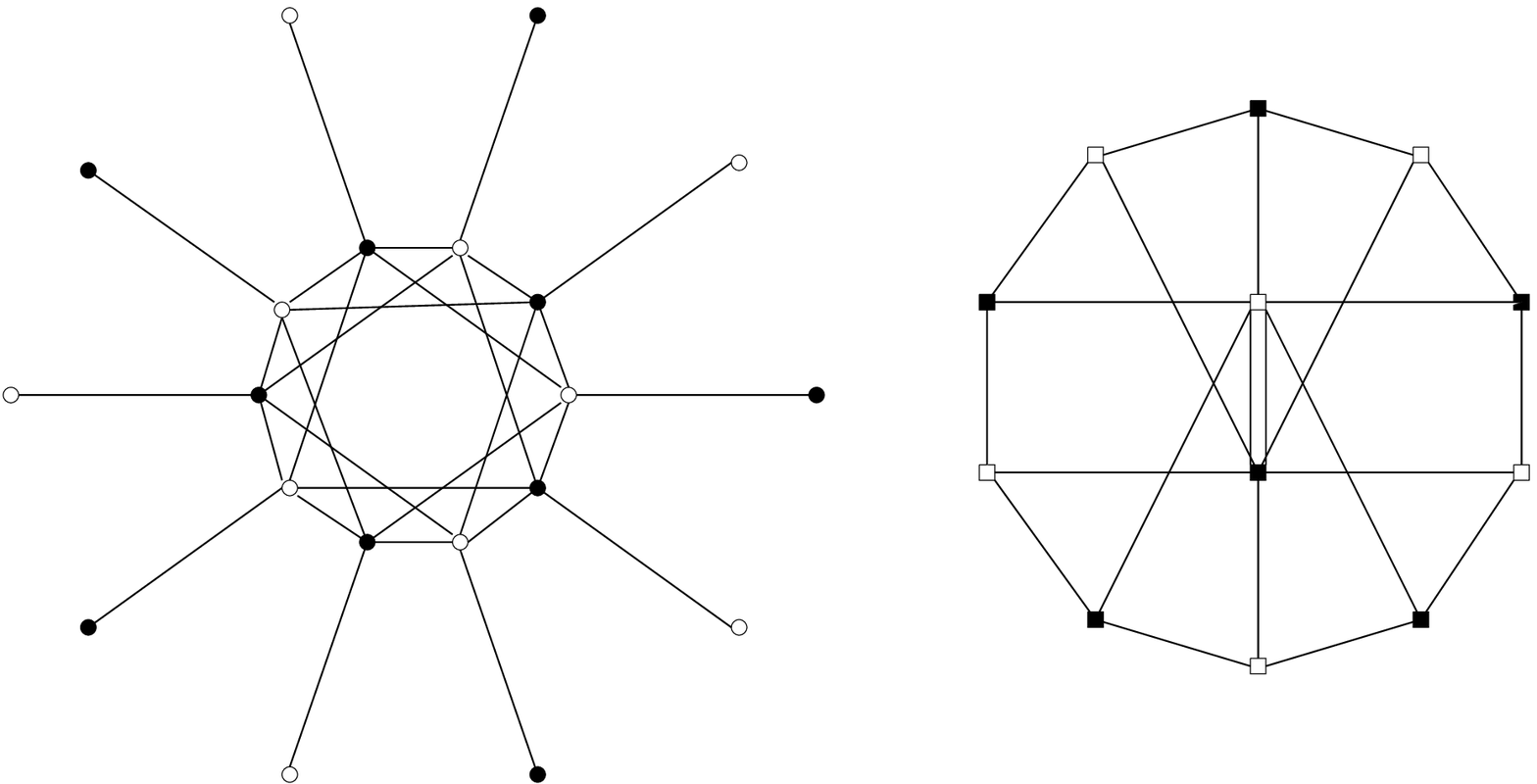}{12cm}\figlabel\geetwo

We conclude with the remark that an example from the last series
in  (7) (see Appendix A of [5]) suggests that the
above scheme might require some modifications -- either extending
the meaning of the graphs, allowing for noninteger values of the
matrix elements of the adjacency matrices, or, the meaning of the
$N$ - algebra, allowing for a  noncommutative extension (see also
[10]). 
\bigskip
 
{\bf Acknowledgments}
\medskip

V.B.P.  was partially supported by the Bulgarian Foundation for
Scientific Research under contract Ph-643.
\medskip

{\bf References}
\medskip

\item{[1]} A. Cappelli, C. Itzykson and J.-B. Zuber,
 Nucl. Phys. {\bf B280} [FS18] ,
445 (1987),  Comm. Math. Phys. {\bf 113} (1987) 1;

 A. Kato,  Mod. Phys. Lett.  {\bf A2}, 585 (1987) .

\item{[2]} V. Pasquier,  Nucl. Phys.  {\bf B285} [FS19], 
162 (1987),  J. Phys.  {\bf A20}, 5707 (1987).

\item{[3]} I.K. Kostov,  Nucl. Phys.  {\bf B 300} [FS22],
559 (1988). 

\item{[4]} P. Di Francesco and J.-B. Zuber,
 Nucl. Phys.  {\bf B338}, 602 (1990), and in  "Recent
Developments in Conformal Field Theories", Trieste Conference  
1989, S. Randjbar-Daemi, E. Sezgin and J.-B. Zuber eds., World
Scientific  1990; 

P. Di Francesco,  Int. J. Mod. Phys.  {\bf A7},  407 (1992).

\item{[5]} V.B. Petkova and J.-B. Zuber,  Nucl. Phys.
{\bf B438}, 347 (1995); ibid {\bf B463}, 161 (1996).

\item{[6]} E. Bannai and T. Ito,  "Algebraic Combinatorics I:
Association Schemes", ~ Benjamin/Cummings (1984).

\item{[7]} F.A. Bais and P.G. Bouwknegt,  Nucl. Phys.  {\bf
B279}, 561 (1987);  

A.N. Schellekens and N.P. Warner,  Phys. Rev. D  {\bf
34}, 3092 (1986).

\item{[8]} J. Fuchs, A.N. Schellekens and C. Schweigert,
 Nucl. Phys.   {\bf  B473}, 323 (1996); 

 E. Baver and D. Gepner, 
 Mod.Phys. Lett.  {\bf A11}, 1929 (1996).

\item{[9]} A. Ocneanu, communication at the Workshop "Low
Dimensional Topology, Statistical Mechanics and Quantum Field
Theory", Fields Institute, Waterloo, Ontario, April 26--30, 1995.

\item{[10]} Feng Xu,  New braded endomorphisms from conformal
inclusions, preprint (1996).

\item{[11]} A.N. Schellekens and S. Yankielowicz,  Nucl. Phys.
  {\bf  B327}, 673 (1989).

\bye